\documentclass[letter]{./templates/IEEEtran}

\newcommand{\pdftitle}{Enabling Deep Reinforcement Learning on Energy Constrained Devices at the Edge of the Network}
\usepackage[nolist,nohyperlinks]{acronym}
\usepackage{amsmath}
\usepackage{amssymb} 
\usepackage{breqn}
\usepackage{balance}
\usepackage{cite}
\usepackage{framed}
\usepackage{siunitx}
\usepackage{soul}
\usepackage{multirow}
\usepackage{etoolbox}
\usepackage{standalone}
\usepackage{textcomp}
\usepackage{multirow,booktabs}
\usepackage{amsmath}
\usepackage{amssymb}
\usepackage{bm}
\usepackage{mathtools} 
\usepackage{optidef} 
\usepackage{algorithm}
\usepackage{algpseudocode}
\usepackage{xcolor, colortbl, collcell}
\colorlet{activecolour}{black}

\usepackage{graphicx}
\usepackage{subcaption}
\usepackage[font=small]{caption}

\graphicspath{{./images/}}

\usepackage{tikz}
\usepackage{pgfplots}
\usepgfplotslibrary{groupplots}

\usepackage{url}
\usepackage[bookmarks=false]{hyperref}
\hypersetup{
  backref,
  hidelinks,
  colorlinks=false,
  breaklinks=true,
  bookmarksopen=false,
  pdftitle=\pdftitle,
  pdfauthor={Author}
}

\newtoggle{authorcomment}
\toggletrue{authorcomment}

\definecolor{LightCyan}{rgb}{0.88,1,1}
\definecolor{Gray}{gray}{0.85}

\usetikzlibrary{arrows,decorations.pathmorphing,backgrounds,positioning,fit,petri}
\usetikzlibrary{shapes.misc}
\tikzset{cross/.style={cross out, draw=black, minimum size=2*(#1-\pgflinewidth), inner sep=0pt, outer sep=0pt},
	cross/.default={1pt}}
\usetikzlibrary{spy,backgrounds,patterns}

\newcommand\copyrighttext{%
  \footnotesize \textcopyright 2022 IEEE. Personal use of this material is permitted.
  Permission from IEEE must be obtained for all other uses, in any current or future
  media, including reprinting/republishing this material for advertising or promotional
  purposes, creating new collective works, for resale or redistribution to servers or
  lists, or reuse of any copyrighted component of this work in other works.
  }
\newcommand\copyrightnotice{%
\begin{tikzpicture}[remember picture,overlay]
\node[anchor=north,yshift=-10pt] at (current page.north) {\fbox{\parbox{\dimexpr\textwidth-\fboxsep-\fboxrule\relax}{\copyrighttext}}};
\end{tikzpicture}%
}

\begin{document}
\bstctlcite{IEEEexample:BSTcontrol}

\title{\pdftitle}

\author{\IEEEauthorblockN{Jernej Hribar, and Ivana Dusparic}

\author{\IEEEauthorblockN{Jernej Hribar and Ivana Dusparic \\
}
\IEEEauthorblockA{
CONNECT-Trinity College Dublin, Ireland \\
Emails: \{jhribar, duspari\}@tcd.ie 
}}


}

\maketitle

\begin{acronym}[MACHU]
  \acro{iot}[IoT]{Internet of Things}
  \acro{cr}[CR]{Cognitive Radio}
  \acro{ofdm}[OFDM]{orthogonal frequency-division multiplexing}
  \acro{ofdma}[OFDMA]{orthogonal frequency-division multiple access}
  \acro{scfdma}[SC-FDMA]{single carrier frequency division multiple access}
  \acro{rbi}[RBI]{ Research Brazil Ireland}
  \acro{rfic}[RFIC]{radio frequency integrated circuit}
  \acro{sdr}[SDR]{Software Defined Radio}
  \acro{sdn}[SDN]{Software Defined Networking}
  \acro{su}[SU]{Secondary User}
  \acro{ra}[RA]{Resource Allocation}
  \acro{qos}[QoS]{quality of service}
  \acro{usrp}[USRP]{Universal Software Radio Peripheral}
  \acro{mno}[MNO]{Mobile Network Operator}
  \acro{mnos}[MNOs]{Mobile Network Operators}
  \acro{gsm}[GSM]{Global System for Mobile communications}
  \acro{tdma}[TDMA]{Time-Division Multiple Access}
  \acro{fdma}[FDMA]{Frequency-Division Multiple Access}
  \acro{gprs}[GPRS]{General Packet Radio Service}
  \acro{msc}[MSC]{Mobile Switching Centre}
  \acro{bsc}[BSC]{Base Station Controller}
  \acro{umts}[UMTS]{universal mobile telecommunications system}
  \acro{Wcdma}[WCDMA]{Wide-band code division multiple access}
  \acro{wcdma}[WCDMA]{wide-band code division multiple access}
  \acro{cdma}[CDMA]{code division multiple access}
  \acro{lte}[LTE]{Long Term Evolution}
  \acro{papr}[PAPR]{peak-to-average power rating}
  \acro{hn}[HetNet]{heterogeneous networks}
  \acro{phy}[PHY]{physical layer}
  \acro{mac}[MAC]{medium access control}
  \acro{amc}[AMC]{adaptive modulation and coding}
  \acro{mimo}[MIMO]{multiple input multiple output}
  \acro{rats}[RATs]{radio access technologies}
  \acro{vni}[VNI]{visual networking index}
  \acro{rbs}[RB]{resource blocks}
  \acro{rb}[RB]{resource block}
  \acro{ue}[UE]{user equipment}
  \acro{cqi}[CQI]{Channel Quality Indicator}
  \acro{hd}[HD]{half-duplex}
  \acro{fd}[FD]{full-duplex}
  \acro{sic}[SIC]{self-interference cancellation}
  \acro{si}[SI]{self-interference}
  \acro{bs}[BS]{base station}
  \acro{fbmc}[FBMC]{Filter Bank Multi-Carrier}
  \acro{ufmc}[UFMC]{Universal Filtered Multi-Carrier}
  \acro{scm}[SCM]{Single Carrier Modulation}
  \acro{isi}[ISI]{inter-symbol interference}
  \acro{ftn}[FTN]{Faster-Than-Nyquist}
  \acro{m2m}[M2M]{machine-to-machine}
  \acro{mtc}[MTC]{machine type communication}
  \acro{mmw}[mmWave]{millimeter wave}
  \acro{bf}[BF]{beamforming}
  \acro{los}[LOS]{line-of-sight}
  \acro{nlos}[NLOS]{non line-of-sight}
  \acro{capex}[CAPEX]{capital expenditure}
  \acro{opex}[OPEX]{operational expenditure}
  \acro{ict}[ICT]{information and communications technology}
  \acro{sp}[SP]{service providers}
  \acro{inp}[InP]{infrastructure providers}
  \acro{mvnp}[MVNP]{mobile virtual network provider}
  \acro{mvno}[MVNO]{mobile virtual network operator}
  \acro{nfv}[NFV]{network function virtualization}
  \acro{vnfs}[VNF]{virtual network functions}
  \acro{cran}[C-RAN]{Cloud Radio Access Network}
  \acro{bbu}[BBU]{baseband unit}
  \acro{bbus}[BBU]{baseband units}
  \acro{rrh}[RRH]{remote radio head}
  \acro{rrhs}[RRH]{Remote radio heads} 
  \acro{sfv}[SFV]{sensor function virtualization}
  \acro{wsn}[WSN]{Wireless Sensor Network} 
  \acro{bio}[BIO]{Bristol is open}
  \acro{vitro}[VITRO]{Virtualized dIstributed plaTfoRms of smart Objects}
  \acro{os}[OS]{operating system}
  \acro{www}[WWW]{world wide web}
  \acro{iotvn}[IoT-VN]{IoT virtual network}
  \acro{mems}[MEMS]{micro electro mechanical system}
  \acro{mec}[MEC]{Mobile edge computing}
  \acro{coap}[CoAP]{Constrained Application Protocol}
  \acro{vsn}[VSN]{Virtual sensor network}
  \acro{rest}[REST]{REpresentational State Transfer}
  \acro{aoi}[AoI]{Age of Information}
  \acro{lora}[LoRa\texttrademark]{Long Range}
  \acro{iot}[IoT]{Internet of Things}
  \acro{snr}[SNR]{Signal-to-Noise Ratio}
  \acro{cps}[CPS]{Cyber-Physical System}
  \acro{uav}[UAV]{Unmanned Aerial Vehicle}
  \acro{rfid}[RFID]{Radio-frequency identification}
  \acro{lpwan}[LPWAN]{Low-Power Wide-Area Network}
  \acro{lgfs}[LGFS]{Last Generated First Served}
  \acro{lmmse}[LMMSE]{Linear Minimum Mean Square Error}
  \acro{rl}[RL]{Reinforcement Learning}
  \acro{nb-iot}[NB-IoT]{Narrowband IoT}
  \acro{lorawan}[LoRaWAN]{Long Range Wide Area Network}
  \acro{mdp}[MDP]{Markov Decision Process}
  \acro{ann}[ANN]{Artificial Neural Network}
  \acro{dqn}[DQN]{Deep Q-Network}
  \acro{mse}[MSE]{Mean Square Error}
  \acro{ml}[ML]{Machine Learning}
  \acro{cpu}[CPU]{Central Processing Unit}
  \acro{ddpg}[DDPG]{Deep Deterministic Policy Gradient}
  \acro{ai}[AI]{Artificial Intelligence}
  \acro{gp}[GP]{Gaussian Processes}
  \acro{drl}[DRL]{Deep Reinforcement Learning}
  \acro{mmse}[MMSE]{Minimum Mean Square Error}
  \acro{fnn}[FNN]{Feedforward Neural Network}
  \acro{eh}[EH]{Energy Harvesting}
  \acro{dl}[DL]{Deep Learning}
\end{acronym}

\begin{abstract}
Deep Reinforcement Learning (DRL) solutions are becoming pervasive at the edge of the network as they enable autonomous decision-making in a dynamic environment. However, to be able to adapt to the ever-changing environment, the DRL solution implemented on an embedded device has to continue to occasionally take exploratory actions even after initial convergence. In other words, the device has to occasionally take random actions and update the value function, i.e., re-train the Artificial Neural Network (ANN), to ensure its performance remains optimal. Unfortunately,  embedded devices often lack processing power and energy required to train the ANN. The energy aspect is particularly challenging when the edge device is powered only by a means of Energy Harvesting (EH). To overcome this problem, we propose a two-part algorithm in which the DRL process is trained at the sink. Then the weights of the fully trained underlying ANN are periodically transferred to the EH-powered embedded device taking actions. Using an EH-powered sensor, real-world measurements dataset, and optimizing for Age of Information (AoI) metric, we demonstrate that such a DRL solution can operate without any degradation in the performance, with only a few ANN updates per day.

\end{abstract}

\acresetall

\begin{IEEEkeywords}
Age of Information, Edge computing, Energy Harvesting, Deep Reinforcement Learning 
\end{IEEEkeywords}

\copyrightnotice

\section{Introduction}
\label{sec:intro}

Recently, we have witnessed a rise in \ac{ann} applications at the edge of the network. In most cases, the \ac{ann} is trained using a large quantity of data on a high-performance computing system, and only a fully trained \ac{ann} is executed on the resource-constrained embedded device at the edge. This approach is highly effective in a supervised learning setting as the device performs only an inference. However, for a \ac{drl}-based solution, which combines the \ac{rl} with  \ac{ann}, the \ac{ann} has to be regularly updated due to continuous exploring in interaction with the environment, to avoid performance degradation. 

Training an \ac{ann} on an embedded device is often challenging due to a lack of memory, processing power, and energy. The latter is especially problematic for embedded edge devices, as the energy used for training the \ac{ann} is enormous for such devices~\cite{garcia2019estimation}, e.g., back-propagation for the entire batch of experiences in a \ac{dqn} algorithm~\cite{mnih2013playing}. Consequently, training on a constrained embedded device is practically impossible. An alternative solution is to perform training on another more resource-rich device, e.g., cloudlet server~\cite{chen2019deep}, and periodically transmit the \ac{ann} to the embedded device. In other words, the embedded device will make decisions and collect information, but the processing-intense task of training the \ac{ann} will be carried out by the unconstrained device. However, transmitting updated \ac{ann} weights to the embedded device also requires substantial energy. In this paper, we explore if an embedded device can support a \ac{drl}-based solution without performing training and rely only on a sporadic update of \ac{ann} weight values.


To test the feasibility of our approach, we focus on a system with a single sensor transmitting measurements to the sink, and measure the performance of the system using the \ac{aoi} metric~\cite{abd2019role}. The \ac{aoi} measures the timeliness of collected information and is becoming ever more important in the \ac{iot} network\cite{abd2019role, yates2015lazy, wu2017optimal, hribar2018using}. By measuring the time passed since a source of information, e.g., a sensor, has generated new information in the form of a status update, e.g., a measurement, it is possible to assess the value of the collected information at the sink, e.g., monitor. The underlying assumption is that the more recent the received information is, the more relevant it is for the system. However, when the embedded device has limited energy available, e.g., a non-rechargeable battery or a reliance on \ac{eh}, there is a fundamental trade-off between the timeliness of collected information and the energy consumption~\cite{yates2015lazy}. In such a system, the device has to balance the frequency with which it transmits status updates and the available energy. In our work, we focus on a system in which a source of information can replenish its energy through \ac{eh}; therefore finding the optimal update policy that minimises the average \ac{aoi} is not trivial. Often, the optimal policy assumes that the amount of energy the \ac{eh}-powered source will collect~\cite{wu2017optimal, hribar2018using} is known to the system. However, typically, such information is not available as the collected energy varies over time. Therefore, solutions capable of adapting to the dynamic environment are more suitable, as considered in our work.


In this paper, we are the first to propose a method that enables the adoption of \ac{drl} on energy-constrained devices. Our solution is a two-part algorithm. The first part is deployed on an energy-constrained device, making decisions and collecting information. The second part of the algorithm is on the resource-unconstrained device performing energy-intense training. The weights of the fully trained \ac{ann}, crucial to the correct performance of \ac{drl}, are periodically transmitted to the device. However, each transfer of \ac{ann} weights consumes a high amount of the embedded device's energy. Therefore, we also balance the frequency of transmitting \ac{ann} weights on top of finding the optimal policy for transmitting status updates. We show that our approach enables the constrained device to learn an updating policy that minimises the average \ac{aoi} in the system of the \ac{eh}-powered device. To provide realism, we use real illuminance measurements to model the time-varying energy arrival over ten days for a device relying on a solar panel to harvest energy. Additionally, we demonstrate that the proposed approach performance is comparable to an unconstrained \ac{drl} solution, i.e., an approach that ignores all device constraints and is able to perform all actions, including the training process, on the device itself. Finally, we demonstrate that our solution can outperform heuristic updating policies under three different energy profiles.


\section{Related Work}
\label{sec:related}

In the literature, there are three prevailing methods of employing \ac{rl} algorithms on energy-constrained devices. The first method is executing a computationally light \ac{rl} algorithm directly on constrained devices~\cite{masadeh2018reinforcement, aoudia2018rlman}. For example, the authors in~\cite{masadeh2018reinforcement} employed simple State–action–reward–state–action (SARSA) and $\epsilon$-greedy algorithms to maximise the sensor's throughput by controlling its transmission power and considering available energy. While the authors in~\cite{aoudia2018rlman} used an actor-critic method to manage power on a \ac{eh}-powered sensor to maximise the quality of service. Unfortunately, the limitations of using such \ac{rl} algorithms are their inability to perform well in a system with a large state space, as is the case in many real-world scenarios. Usually, an \ac{ann} with multiple hidden layers is used as a function approximation for the value function. Consequently, the second prevailing method is implementing a \ac{drl} algorithm on another resource unconstrained device that manages the energy-constrained device with control messages~\cite{globecom2021, chu2018reinforcement, hribar2020energy}. For example, \ac{drl}-based energy-aware scheduling schemes, implemented on a gateway, for battery-powered or \ac{eh}-powered devices were proposed in~\cite{globecom2021, chu2018reinforcement, hribar2020energy}. At the same time, the third method was developed in which the \ac{ann} is trained in an offline manner and then deployed on the constrained device~\cite{sharma2019distributed, li2019deep}.  In such a case, the constrained device exploits the pre-trained \ac{ann} to minimise training carried out on the constrained device. For example, such a scheme was proposed in~\cite{li2019deep} to adjust the low-power sensors' transmission power dynamically. The work in~\cite{sharma2019distributed} proposed a design of a distributed \ac{drl} solution for controlling the device's power. In the latter, authors employed offline learning to mitigate the energy consumption of embedded devices. 
In contrast to these three methods, we enable the full \textit{deep} \ac{rl}, i.e., \ac{rl} with value function approximated by the \ac{ann}, on energy-constrained embedded devices. Our solution relies on continuous updating rather than once-off transfer, with heavy training at each step offloaded to a more resource-rich device.

\section{System Model}
\label{sec:related}

\begin{figure}
	\centering
	\includegraphics[width=3.3in]{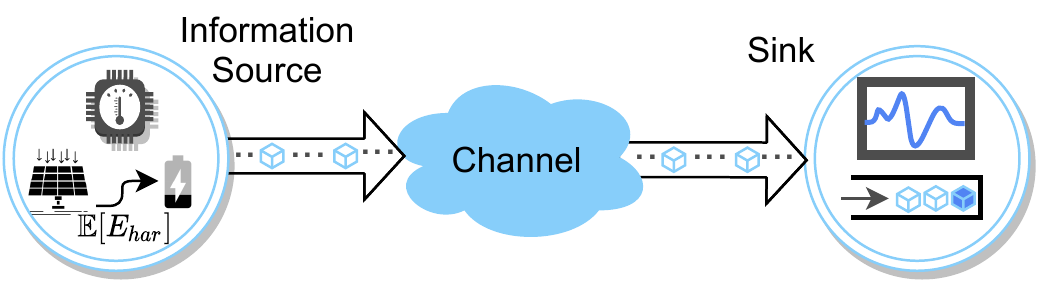}
	\caption{System model illustration.}
	\label{fig:system_model}
	\vspace{-10pt}
\end{figure}

We consider a sensor network consisting of one wireless sensor node, i.e., information source, and one sink, e.g., monitor, as shown in Fig.~\ref{fig:system_model}. 
We assume that the source operates using sleep and wake-up cycles, a commonly employed practice to preserve low-power sensors' energy. For simplicity, we consider a slotted time $t \in \mathcal{T} = \{1, \hdots, T\} $. In each time step, the source consumes $E_{M}$ amount of energy to support the essential operations of the source and to take the observation $O_t$, e.g., temperature measurement. 

The source node harvests the energy from, e.g., light, vibrations, or wind. The harvested energy is proportional to the current $I_{EH}(t)$ and voltage $U_{EH}(t)$ originating from the \ac{eh} unit. The source stores the collected energy in a capacitor of finite-size $B$. Consequently, the source's energy $E(t)$ is limited to an interval; $E(t) \in [0, B]$. The $E(t)$ varies over time due to the variable $I_{EH}(t)$ and due to the energy cost $E_{TR}(t)$  of transmitting a status update.

\subsection{Timeliness of Collected Information}

The source generates a status update $U_t$ at time step $t$ if the source decides to transmit new information to the sink for collection. Hence, we can define the process $\Delta(t)$, representing the timeliness of collected information, i.e., status updates, as follows:

\begin{equation}\label{eq:delta1}
\Delta(t)\coloneqq t-G(t).
\end{equation}

\noindent where $G(t)$ represents a time instance in which a status update is generated. We assume that the status update $U_t$ also contains observations the source has obtained in the previous time steps, as is often the case in real deployments. The source employs buffer $\mathcal{B}$ with length $M$ to save past observations. Meaning, that the newly generated status update contains a set of multiple observations, i.e., $U_t = \{ O_{t}, O_{t-1},\hdots O_{t-M-1} \}$.


The first metric we employ to measure the timeliness of information in the system is average \ac{aoi}. In discrete systems it is defined as follows:


\begin{equation}\label{eq:average_age}
\overline{\Delta}(t) = \frac{1}{T} \lim_{T \rightarrow \inf} \sum_{t=1}^{T} \Delta(t).
\end{equation}

\noindent The average value enables us to assess how the system performs as a whole in terms of \ac{aoi}. However, the main limitation of the average \ac{aoi} metric is that it is not possible to determine how are the \ac{aoi} values of status updates spread around the average. To that end, the \emph{peak} \ac{aoi} was proposed by Costa et. al.~\cite{costa2016age} to provide a better characterization of a process  $\Delta(t)$. First, we define $P_i$ as a peak, i.e., the highest value of the process $\Delta(t)$ before a new status updated is received at the sink. Then, the peak \ac{aoi} is determined as:

\begin{equation}\label{eq:peak_age} 
\Delta_{peak}(t) = \frac{1}{K} \sum_{i=1}^{K} P_i,
\end{equation}

\noindent where $K$ represents the number of received status updates. In \ac{eh}-powered sources, the peak \ac{aoi} is connected with down-time, i.e., $T_{down}$. We define $T_{down}$ as a time during which the source does not have enough energy to transmit a new status update, i.e., $E(t) < E_{TR}(t)$. In general, if a source experiences high $T_{down}$, the peak \ac{aoi} is higher, a behaviour we analyse in more depth in the validation Section \ref{sec:validation}.

\subsection{Minimising AoI in an Energy-constrained System}

The system's main goal is to find a status updating policy $\pi$ that will ensure that the \ac{eh}-powered source will minimise the average and peak \ac{aoi}. Therefore, we formulate the problem the system is facing as:

\begin{equation}
\begin{aligned}
\min_{\pi}  \quad &  \overline{\Delta}(t), \Delta_{peak}(t)\\
\textrm{s.t.}  \quad & E(t)  \geq 0, \forall \: t=1,\ldots,T, \\
& T_{down} = 0, \forall \: t=1,\ldots,T.
\end{aligned}
\label{eq:problem_formulation}
\end{equation}

\noindent Due to the limited and randomly time-varying nature of $E(t)$, deriving an updating policy that will minimise the average \and peak \ac{aoi} in the \ac{eh}-powered system is a non-trivial task. To that end, we employ \ac{drl} and modify it to be able to overcome hardware constraints of a low-cost \ac{eh}-powered source, as we describe in the next section.

\section{Enabling DRL on Energy Constrained Devices}
\label{sec:related}

In each time step, the source has to decide whether to transmit a status update or wait. Each option has a different energy cost. If the source decides to transmit, it uses the energy required to transmit a status update; if it decides to wait, the value of the \ac{aoi} will increase. Therefore, it is paramount that the source finds a policy capable of preserving the energy while also ensuring that the resulting average \ac{aoi} will be minimal. To that end, we employ a \ac{dqn} algorithm~\cite{mnih2013playing} to enable the source to decide dynamically. Alternatively, any other \ac{drl} algorithm, e.g., \ac{ddpg}~\cite{lillicrap2015continuous}, Soft Actor-Critic (SAC)~\cite{haarnoja2018soft}, etc., could be applied with a minimal change to our design. Utilizing the \ac{ann} as a function approximation enables us to use real positive values, as obtained by the source, as an input to the decision-making process.

\subsection{States, Actions, and Reward Function}

We define states, actions, and rewards for our proposed DRL-based approach as follows: 

\noindent \textit{State:}
The state $s_t$ consist of the three most critical parameters for the source's decision: the amount of energy currently available on the source, \ac{aoi} value, and the harvesting current, i.e., $s_{t} = (E(t), \Delta(t), I_{EH}(t)$. Due to the high variability in the values of the parameters in the state vector, the number of possible states results in millions, motivating the use of DRL.

\noindent  \textit{Action:} The source only has two actions available: transmit or do not transmit. Therefore, the action value ($a_t \in \mathcal{A}$) can only take two different values, i.e., $\mathcal{A} = \{0,1\}$.

\noindent \textit{Reward:} The reward $r_t$ consist of two parts, i.e., $r_t(t) = r_a(t) + r_e(t)$. The first part of the reward depends on the action and average \ac{aoi} achieved in the past day. When the source decides to not transmit a new status update the reward is proportional to one tenth of the average \ac{aoi}, i.e., $r_a(t) = 1 - \frac{1}{10}\overline{\Delta}(t)$. When the source transmits a new status update, the reward is much higher. However. to ensure stability we had to limit the reward as follows:

\begin{equation}
r_a(t)= 
 \begin{cases} 
2.5 (40 - \overline{\Delta}(t)) + 2 - \frac{1}{10}\overline{\Delta}(t), &  \operatorname{if} \overline{\Delta}(t) \leq 40 \\

\phantom{AAAA} 2 - \frac{1}{10}\overline{\Delta}(t), & \operatorname{otherwise}\\

\end{cases}
\textbf{\label{acc_rewd}}
\end{equation}

\noindent The energy part of the reward function activates once the available energy in the source is below the threshold, i.e., $r_e(t) = - 1000; \operatorname{ if }E(t) \leq 0.15 B$. 

\subsection{Two-part DQN for Energy Constrained Devices}

We describe the proposed two-part algorithm in Algorithms 1 and 2. Algorithm 1 is executed on the \ac{eh}-powered source, while Algorithm 2 runs on the sink. At the start, the \acp{ann}, representing the Q-value function, on both the source and the sink are initialised with identical random weights($\theta_{c}, \theta_{p}$, and $\theta_{t}$). In practice, this can be easily achieved using the same random seed number.

\begin{algorithm}
 \caption{The first part of the proposed algorithm running on the energy-constrained device (source).}
 \begin{algorithmic}[1]
    \State Init action-value function $Q_{c}$ with weights $\theta_{c}=\theta_{r}$
    \For{$t=1,T$}
        \If{$E(t)>E_M$}
        \State Take observation $O_t$ and store it to buffer $\mathcal{B}$ 
        \State With probability $\epsilon$ select a random action $a_t$
        \State Otherwise select $a_t$ as $argmax_a Q_{c}(s_t,a | \theta_{c}) $
        \If{$a_t == True$ \textbf{and}  $E(t) \geq E_{TR}(t)$}
        \State Transmit $U_t$ to the sink for collection
        \EndIf
        \If{$T_u \geq T_{ANN}$ \textbf{and} $E(t) \geq E_{ANN}(t)$}
        \State Update $Q_{c}$ using $Q_p$ weights, i.e., $\theta_{c}=\theta_{t}$
        \State $T_{u} \gets 0$ \Comment{Set ANN update timer to zero.}
        \Else
        \State $T_{u} \gets T_{u} + 1$ 
        \EndIf
        \EndIf
    \EndFor
 \end{algorithmic}
\end{algorithm}

\textit{Algorithm 1}: At the start of each time step, the \ac{eh}-powered source wakes up, takes the new measurement $O_t$ and adds it to buffer $\mathcal{L}$. In the next step, the source will, with probability $\epsilon$, select the action randomly or use cached \ac{ann} to decide on the action, i.e., $argmax_a Q_{c}(s_t,a | \theta_{c})$. The source will send a new status update, containing up to $M$ observations, to the sink, provided that the source has enough energy to make the transmission. Before entering sleep mode again, the source checks if the weights $Q_c$ should be updated. The $T_{ANN}$ represents the minimum number of time steps that need to pass before the \ac{ann} is updated. While $T_u$ counts time passed since the last update.

 \begin{algorithm}
 \caption{The second part of the proposed algorithm running on the resource unconstrained device (sink).}
 \begin{algorithmic}[1]
    \item Init policy action-value function $Q_{p}$ with weights $\theta_{p}=\theta_{r}$ 
    \item Init target action-value function $Q_{t}$ with weights $\theta_{t} = \theta_{p}$ 
   \item Init replay memory $\mathcal{D}$ to capacity $D$
   \For{$t=1,T$} 
        \If{sink receives $U_t$}
        \For{$O_i$ in $U_t$}
        \State Parse $O_i$ to get $s_t, s_{t+1}, r_t, a_t$ and store in $\mathcal{D}$
        \EndFor
        \EndIf
        \State Sample random batch of $N$ experiences from $\mathcal{D}$
        \For{every experience $\{s_j, s_{j+1}, r_j, a_j\}$ in batch}
        \State Set $y_j = r_j + \gamma max_{a_{j+1}} Q_{t}(s_{j+1}, a_{j+1})$
        \EndFor
        \State  Calculate the loss  $\mathcal{L} = \frac{1}{N} \sum_{j=0}^{N-1} (Q(s_j, a_j) - y_j )^2$
        \State Update $Q_p$ by minimising the loss  $\mathcal{L}$
        \State Every $C$ steps, update $Q_t$ using $Q_p$, i.e, $\theta_{t} = \theta_{p}$ 
    \EndFor
 \end{algorithmic} 
 \end{algorithm} 

\textit{Algorithm 2}: On the other hand, the sink will parse the received status update $U_t$ and determine the reward for every received observation. Thus, the sink can reconstruct up to $M$ past episodes necessary to obtain the \emph{experience}   (i.e., a tuple consisting of a state, action, reward, and future state, $\{s_t, s_{t+1}, r_t, a_t\} $), essential for the training of any \ac{drl} algorithm, where $M$ denotes the number overall measurements the device transmits. These experiences are then used to train the policy \ac{ann} using batch learning. The learning process, i.e., calculating the loss $\mathcal{L}$ and updating policy \ac{ann} $Q_p$ by minimising the loss $\mathcal{L}$, follows the standard \ac{dqn} algorithm approach as described in~\cite{mnih2013playing}. 

 \begin{figure}
	\centering
	\includegraphics[width=3.5in]{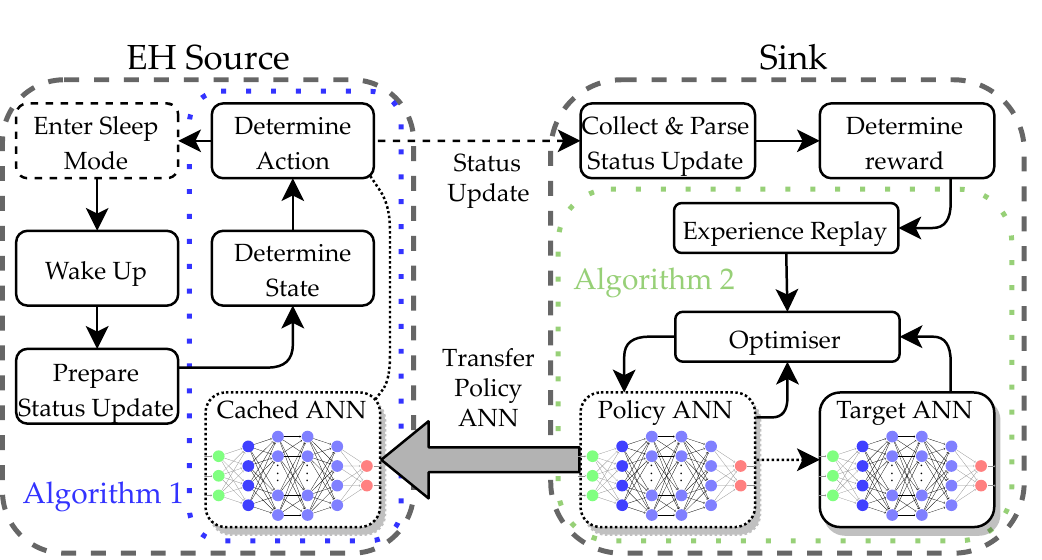}
\caption{The block diagram of the proposed solution. }
	\label{fig:solution}
	\vspace{-15pt}
\end{figure}

Fig.~\ref{fig:solution} depicts the block diagram overview of the proposed solution and illustrates the interactions between the energy-constrained device and the sink (resource unconstrained device). 

\subsection{Implementation}

We employ \ac{dqn} structure with four hidden layers. The first and the last hidden layer consist of four neurons, while the inner two hidden layers are implemented with eight neurons each. All layers use the ReLU activation function. To prevent over-fitting, we have added a ten per cent dropout layer between the hidden layers, and we also applied batch normalisation before the first hidden layer. We list the rest of the \ac{dqn} algorithm hyperparameters in Table \ref{table_hyper_param}. 

\begin{table}[ht]
	\centering
	\caption{DQN Hyperparameters}
	\label{table_hyper_param}
	\begin{tabular}{ll || ll}
		\toprule
		\begin{tabular}[c]{@{}c@{}} Hyperparameter $\phantom{AAAA}$ \end{tabular} & 
		\begin{tabular}[c]{@{}c@{}} Value \end{tabular} &
		\begin{tabular}[c]{@{}c@{}} Hyperparameter $\phantom{AAAA}$\end{tabular} & 
		\begin{tabular}[c]{@{}c@{}} Value \end{tabular} \\
		\midrule 
		
		\begin{tabular}[c]{@{}c@{}}  Batch size $N$  \end{tabular} & 
		\begin{tabular}[c]{@{}c@{}}   $64$ \end{tabular} &
		
		\begin{tabular}[c]{@{}c@{}}  Memory size $D$ \end{tabular} & 
		\begin{tabular}[c]{@{}c@{}}   $10^5$\end{tabular} \\

		\begin{tabular}[c]{@{}c@{}}  Optimizer  \end{tabular} & 
		\begin{tabular}[c]{@{}c@{}}   Adam \end{tabular} &
		\begin{tabular}[c]{@{}c@{}}  Loss Function  \end{tabular} & 
		\begin{tabular}[c]{@{}c@{}}   MSE \end{tabular} \\

		
		
		\bottomrule   
\end{tabular}
	\vspace{-15pt}
\end{table}

\section{Validation}
\label{sec:validation}

\begin{figure}
	\centering

\begin{tikzpicture}
\begin{groupplot}[group style={group name=my_plots, group size=1 by 2,vertical sep= 1.25cm}, width=0.75\textwidth, height=0.25\textwidth,]]






\nextgroupplot[
axis lines=middle,
title = (a) Average AoI,
xlabel=$t$ (days),
ylabel=Average AoI(min),
ytick ={600, 1200, 1800, 2400, 3000, 3600, 4200},
yticklabels={10,20,30,40,50,60,70},
xtick={3,5,7,9,11, 13, 15, 17},
xticklabels={1,3,5,7,9,11, 13, 15},
 x label style={at={(current axis.right of origin)},anchor=north, below=0mm, xshift=-3mm},
legend style={at={(1,1)}},
xmin= 2,xmax=19,
ymin= 0,ymax= 5000,
every tick/.style={
	black,
	semithick,
}]
\draw[black!60,dashed] (0,420) -- (435, 420);
\draw[black!60,dashed] (0,360) -- (435, 360);
\draw[black!60,dashed] (0,300) -- (435, 300);
\draw[black!60,dashed] (0,240) -- (435, 240);
\draw[black!60,dashed] (0,180) -- (435, 180);
\draw[black!60,dashed] (0,120) -- (435, 120);
\draw[black!60,dashed] (0,60) -- (125, 60);

\addplot[style=loosely dashed, mark size=2pt, line width=1pt, color=red!95, mark=triangle*, smooth, mark options={solid}]
table[x=days,y=avg_aoi,col sep=comma]{csv_data/basic_results.csv};


\addplot[style=loosely dashed, mark size=2pt, line width=1pt, color=red!75, mark=square*, smooth, mark options={solid}]
table[x=days,y=avg_aoi,col sep=comma]{csv_data/basic_results_2.csv};

\addplot[style=loosely dashed, mark size=2pt, line width=1pt, color=red!55, mark=pentagon*, smooth, mark options={solid}]
table[x=days,y=avg_aoi,col sep=comma]{csv_data/basic_results_3.csv};


\addlegendentry{1-a-day-transfer};
\addlegendentry{2-a-day-transfer};
\addlegendentry{3-a-day-transfer};




\nextgroupplot[
axis lines=middle,
title = (b) Down-time,
xlabel=$t$ (days),
ylabel= $T_{down}$ (h),
ytick ={600, 1200, 1800, 2400, 3000, 3600},
yticklabels={1,2,3,4,5,6},
xtick={3,5,7,9,11, 13, 15, 17},
xticklabels={1,3,5,7,9,11, 13, 15},
 x label style={at={(current axis.right of origin)},anchor=north, below=0mm, xshift=-3mm},
legend style={at={(1,1)}},
xmin= 2,xmax=19,
ymin= 0,ymax= 4750,
every tick/.style={
	black,
	semithick,
}]

\draw[black!60,dashed] (0,360) -- (435, 360);
\draw[black!60,dashed] (0,300) -- (435, 300);
\draw[black!60,dashed] (0,240) -- (435, 240);
\draw[black!60,dashed] (0,180) -- (435, 180);
\draw[black!60,dashed] (0,120) -- (435, 120);
\draw[black!60,dashed] (0,60) -- (435, 60);

\addplot[style=loosely dashed, mark size=2pt, line width=1pt, color=red!95, mark=triangle*, smooth, mark options={solid}]
table[x=days,y=down_time,col sep=comma]{csv_data/basic_results.csv};


\addplot[style=loosely dashed, mark size=2pt, line width=1pt, color=red!75, mark=square*, smooth, mark options={solid}]
table[x=days,y=down_time,col sep=comma]{csv_data/basic_results_2.csv};

\addplot[style=loosely dashed, mark size=2pt, line width=1pt, color=red!55, mark=pentagon*, smooth, mark options={solid}]
table[x=days,y=down_time,col sep=comma]{csv_data/basic_results_3.csv};

\addlegendentry{1-a-day-transfer};
\addlegendentry{2-a-day-transfer};
\addlegendentry{3-a-day-transfer};

\end{groupplot}
\end{tikzpicture}
	\caption{Daily average AoI and down-time.}
	\label{fig:training}
	\vspace{-15pt}
\end{figure}

\begin{figure*}
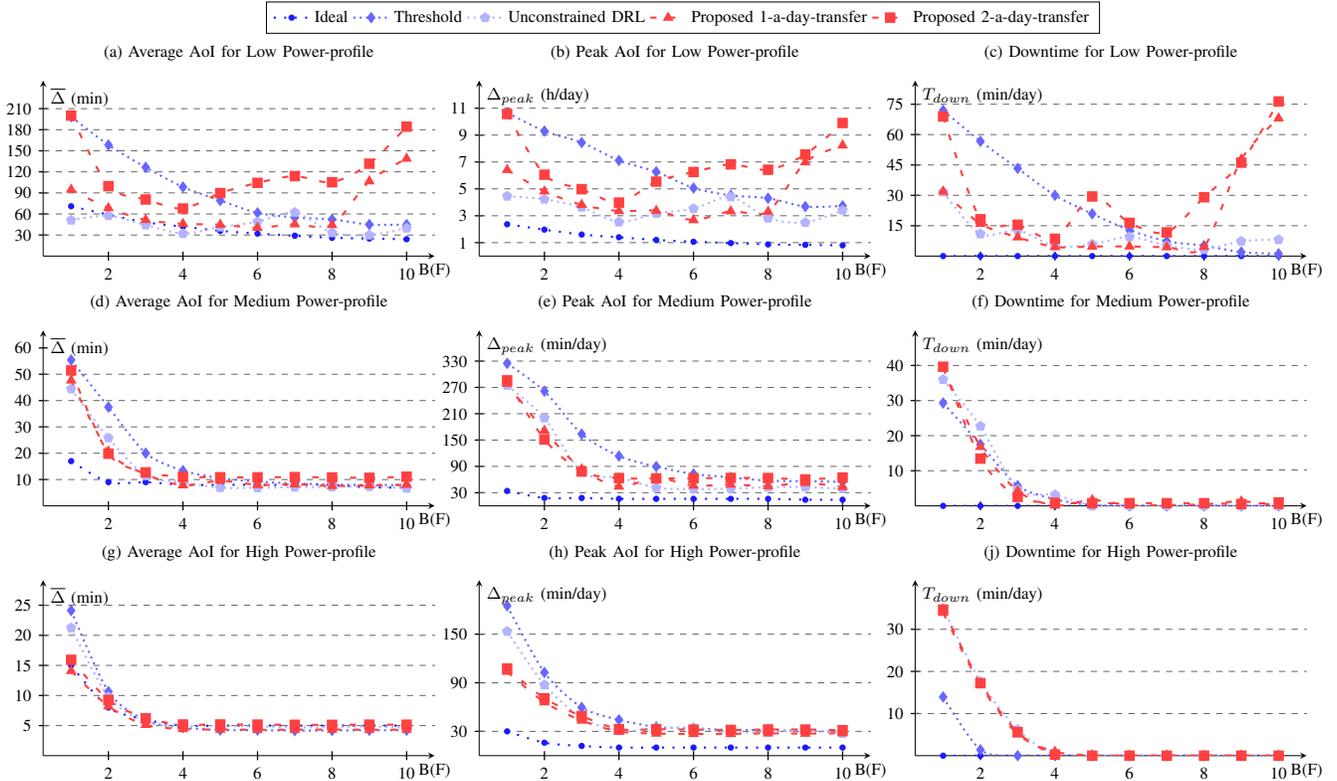

	\centering
	\includestandalone[width=7.0in]{tikz_figures/main_results}
	\caption{Average AoI, peak AoI, and $T_{down}$ for three different power profiles.}
	\label{fig:multiple_sources}
	\vspace{-10pt}
\end{figure*}


In this section, we first show the impact of the frequency of \ac{ann} weight updates on the performance. We continue by validating the feasibility of the proposed mechanism by comparing its performance, in terms of achieved average \ac{aoi}, peak \ac{aoi}, and downtime, to results obtained by three baseline updating policies. To add realism to our simulations, we model the $I_{EH}(t)$ using illuminance measurements obtained in a real sensor deployment at Intel Berkeley Research laboratory~\cite{bodik2004intel}. By combining the illuminance measurements with the measurements performed by Yue et al.~\cite{yue2017development} we can determine the energy an \ac{eh}-powered source collects in a real-world setting. Additionally, we divide the sources based on the amount of energy a source gathers per day into three power profiles: low, medium, and high. Testing the performance under different energy scenarios enables us to demonstrate the proposed solution generalisability. For illustration, a source with a low power profile collects on average $6 J$, a medium one $14 J$, and a high $20 J$ of energy per day.

We model the consumption of the source based on the LoRaWAN class B with spread factor 8\footnote{We rely on the following calculator to determine the energy consumption: \href{https://www.loratools.nl}{www.loratools.nl}.}. The energy required to transmit the status updates varies depending on the number of measurements the source sends. Therefore, the energy for transmitting the status update $E_{TR}$ varies between $35mJ$ to $100mJ$.  In our simulation, the source reports humidity and temperature readings as measured by a sensor in the Intel laboratory deployment. Furthermore,  we assume that the source can successfully transmit a new status update with probability $\eta =0.9$. In our implementation, the \ac{ann}'s weights file size is around $10kB$. In our work, we assume that the sink transmits the file without any compression. Transferring a large amount of data to an embedded device is not unusual as devices are often updated, e.g., updating firmware on embedded device~\cite{abdelfadeel2020make}. According to our estimates, the source would have to listen for around $30s$ to receive the file. Consequently, the energy cost of updating \ac{ann} $E_{ANN}$ is in the range of $0.9J$. We list the remaining energy parameters in Table \ref{simulation_param}.

\begin{table}[ht]
	\centering
	\caption{Simulation Parameters}
	\label{simulation_param}
	\begin{tabular}{ll || ll || ll}
		\toprule
		\begin{tabular}[c]{@{}c@{}} Parameter\end{tabular} & 
		\begin{tabular}[c]{@{}c@{}} Value \end{tabular} &
		
		\begin{tabular}[c]{@{}c@{}} Parameter\end{tabular} & 
		\begin{tabular}[c]{@{}c@{}} Value \end{tabular} &
		
		\begin{tabular}[c]{@{}c@{}} Parameter\end{tabular} & 
		\begin{tabular}[c]{@{}c@{}} Value \end{tabular} \\
		\midrule 
		
		\begin{tabular}[c]{@{}c@{}}  time-step size \end{tabular} & 
		\begin{tabular}[c]{@{}c@{}}  $120s$ \end{tabular} &
		
		\begin{tabular}[c]{@{}c@{}} $E_{ANN}$\  \end{tabular} & 
		\begin{tabular}[c]{@{}c@{}} $900mJ$ \end{tabular} &
		
		
		\begin{tabular}[c]{@{}c@{}}  $E_{M}$\end{tabular} & 
		\begin{tabular}[c]{@{}c@{}}  $1.5mJ$ \end{tabular} \\
		
		
		\begin{tabular}[c]{@{}c@{}} $U_{EH}(t)$\end{tabular} & 
		\begin{tabular}[c]{@{}c@{}}  $3V$ \end{tabular} &
		
		\begin{tabular}[c]{@{}c@{}}  $\eta $\end{tabular} & 
		\begin{tabular}[c]{@{}c@{}} $0.9$ \end{tabular} &
		
		\begin{tabular}[c]{@{}c@{}}  $M$ \end{tabular} & 
		\begin{tabular}[c]{@{}c@{}}  $4$ \end{tabular} \\
		
		

		\bottomrule   
\end{tabular}
	\vspace{-15pt}
\end{table}

\subsection{The Impact of Frequency of  ANN Weights Updates }

In Fig.~\ref{fig:training}, we show how fast, depending on the number of \ac{ann} weight updates a system can make each day, the device converges to the desired behaviour, i.e., finds a policy that minimises the average \ac{aoi}. At the beginning of our experiment, we randomly initialise weights in the \ac{ann} deployed on the device. The selected energy storage capacity is 4 farads, i.e., $B=4F$, and the device collects on average $14J$ of energy each day. When \ac{ann} weights are updated only once per day, the source requires ten days to find the policy that minimises the average \ac{aoi}. On the other hand, when we update \ac{ann} weights two or three times a day, the system will find the policy in only four days. Note that we achieved the numbers of transmissions per day by setting the $T_{ANN}$. For once-a-day \ac{ann} update, we set $T_{ANN}$ to twenty-two hours, for two updates per day to eight hours, and three-a-day updates to six hours. Such a difference was necessary as the solar-powered source collects more energy in the afternoons. Interestingly, when we update the \ac{ann} weights more than two times per day, we can notice deterioration in the performance in the form of downtime ($T_{down}$) as shown in Fig.~\ref{fig:training}~(b). The deterioration occurs as the energy required to update the  \ac{ann} weights three times a day is too high for the selected \ac{ann} size and the energy the source can collect in a day. Overall, the general behaviour of the proposed solution is as shown in Fig.~\ref{fig:training}. However, the best daily \ac{ann} weight update frequency depends on several factors, including \ac{ann} size and energy storage capacity.

\subsection{Age of Information Performance}

We show the performance in terms of achieved average \ac{aoi} over one week of operation in Fig.~\ref{fig:multiple_sources} for the three power profiles. We define the three updating policies we use as baselines to compare the proposed solution to as:

\begin{enumerate}
    \item \textbf{Unconstrained DRL}: A partially idealised approach which assumes that the source consumes zero energy to train the \ac{ann} and therefore trains directly on the device. 
    \item \textbf{Threshold}: The probability that the source will transmit is proportional to its energy level. For example, if the source has $25\%$ of the battery left, the source has a $25\%$  chance it will decide to transmit.
    \item \textbf{Ideal Uniform}: This is a fully idealised approach as we assume that the source employs an oracle that provides the source with the exact information regarding the energy it will collect in the future. Using such information, the source can adopt an ideal strategy that will result in zero downtime.
\end{enumerate}

The obtained simulated results of average \ac{aoi}, i.e., $\overline{\Delta}$, in Fig.~\ref{fig:multiple_sources} (a), (d), and (g) show that the smaller the storage capacity is and the less energy the source can collect, the more advantageous it is to use \ac{drl}. Additionally, updating \ac{ann} weights only once per day results in performance close to the unconstrained \ac{drl} approach. The only exception is the low power-profile, where noticeable degradation occurs for capacitor sizes of $9$ and $10$ farads. Such degradation happens because the source does not have enough energy to transmit new \ac{ann} weights. Surprisingly, updating weights twice per day yields the worst performance in terms of average \ac{aoi}, mostly due to the device consuming more energy in comparison to other approaches.

In Fig.~\ref{fig:multiple_sources} (b), (e), and (h) we show the performance in terms of \emph{peak} \ac{aoi}. The peak \ac{aoi} provides a better characterisation of a process  $\Delta(t)$ as it focuses on the maximal \ac{aoi} value, i.e., the worst case. Using \ac{drl} results in a lower peak \ac{aoi} in comparison to other approaches, and its performance is close to that of the ideal updating policy. Such a result indicates that a \ac{drl} solution is much better at preserving the device's energy to ensure lower peak \ac{aoi}. We observe similar behaviour in the downtime, which we present in Fig.~\ref{fig:multiple_sources} (c), (f), and (j). In general, the greater the $T_{down}$, the longer is the corresponding peak \ac{aoi}. Interestingly, our \ac{drl} solution will result in a longer downtime than a threshold policy but will have lower peak \ac{aoi}. Such behaviour indicates the \ac{drl} solution's superior ability to distribute the source's status updates throughout the day.

The average and peak \ac{aoi} performance is linked to the number of the device's daily transmissions. For example, focusing on a case in which various updating policies achieve similar \ac{aoi} performance, such as medium power profile for $B=5F$, the source relying on the threshold policy will transmit $220$ status updates each day. On the other hand, a source employing unconstrained \ac{drl} will send $170$ status updates each day, $140$ if it relies on the proposed approach with a single daily \ac{ann} weight update, and $120$ transmissions if it uses two daily updates. However, a source will send only $90$ status updates in an ideal policy. Such a result may come as a surprise, but as stated in \cite{yates2015lazy}, the source of information with limited energy might have to act lazy to achieve the optimal \ac{aoi} performance.


\section{Conclusion}
\label{sec:conclusion}

In this paper, we have demonstrated that it is possible to facilitate a \ac{drl}-based solution on a resource-constrained embedded device powered by \ac{eh}. In the proposed approach, the \ac{drl} agent that is implemented on an \ac{eh}-powered device is only taking actions and sensing the environment. At the same time, the \ac{ann} training is performed on a distant unconstrained device, and the weights of the fully trained \ac{ann} are periodically transmitted to the constrained device. We have evaluated the system's performance using the average and peak \ac{aoi}. We have demonstrated that the proposed periodical sending and updating of \ac{ann} weights is feasible, as the resulting performance can be comparable to the ideal solution. 

In future work, we will explore the advantages that a system can gain by leveraging information from more devices. For example, as training is performed on a centralised unit, experiences from multiple sources could be combined. Moreover, a system could transmit trained \ac{ann} weights to numerous sources at the same time to preserve radio time.

\section*{Acknowledgements}

This work was funded in part by the European Regional Development Fund through the SFI Research Centres Programme under Grant No. $13/RC/2077\_P2$ SFI CONNECT, the SFI-NSFC Partnership Programme Grant Number 17/NSFC/5224, and SFI Enable Grant Number 16/SP/3804.

\balance

\bibliographystyle{./templates/IEEEtran}
\bibliography{IEEEabrv,bibliography}

\end{document}